\documentclass[10pt,conference]{IEEEtran}
\IEEEoverridecommandlockouts
\newcommand{\CLASSINPUTtoptextmargin}{2.4cm}
\usepackage{tabularx}
\usepackage{cite}
\usepackage{amsmath,amssymb,amsfonts}
\usepackage{graphicx}
\usepackage{textcomp}
\usepackage{xcolor,soul,framed}
\usepackage{comment}
\usepackage{multirow}
\usepackage{algpseudocode}
\usepackage[linesnumbered,lined,boxed,commentsnumbered,ruled,longend, noend]{algorithm2e}
\usepackage{fancyhdr}

\SetCommentSty{mycommfont}
\SetKwInput{KwInput}{Input}                
\SetKwInput{KwOutput}{Output}              
\def\BibTeX{{\rm B\kern-.05em{\sc i\kern-.025em b}\kern-.08em
    T\kern-.1667em\lower.7ex\hbox{E}\kern-.125emX}}
\IEEEoverridecommandlockouts\IEEEpubid{\makebox[\columnwidth]{ 979-8-3503-1090-0/23/\$31.00~\copyright~2023 IEEE \hfill} \hspace{\columnsep}\makebox[\columnwidth]{}}

\begin{document}


\title{Blockchain and Deep Learning-Based IDS for Securing SDN-Enabled Industrial IoT Environments
}
\author{
    \IEEEauthorblockN{
        Samira Kamali Poorazad, Chafika Benza\"{i}d, and Tarik Taleb 
    }
    \IEEEauthorblockA{
        Oulu University, Oulu, Finland\\
        \{samira.kamalipoorazad; chafika.benzaid; tarik.taleb\}@oulu.fi}
\vspace{-1.3cm}
}
\maketitle

\fancypagestyle{mahmood}{%
   \fancyhf{} 
   \renewcommand{\headrulewidth}{0pt}
   \fancyhead[C]{979-8-3503-1090-0/23/\$31.00~\copyright~2023 IEEE. Reprinting or republishing this material for the purpose of advertising or promotion, reselling or redistributing to servers or lists, or using any copyrighted component in other works must adhere to IEEE policy. The paper has been accepted for publication at \textbf{GlobeCom 2023}, and DOI will be provided as soon as possible.}
}

\thispagestyle{mahmood} 

\begin{abstract}
The industrial Internet of Things (IIoT) involves the integration of Internet of Things (IoT) technologies into industrial settings. However, given the high sensitivity of the industry to the security of industrial control system networks and IIoT, the use of software-defined networking (SDN) technology can provide improved security and automation of communication processes. Despite this, the architecture of SDN can give rise to various security threats. Therefore, it is of paramount importance to consider the impact of these threats on SDN-based IIoT environments. Unlike previous research, which focused on security in IIoT and SDN architectures separately, we propose an integrated method including two components that work together seamlessly for better detecting and preventing security threats associated with SDN-based IIoT architectures. The two components consist in a convolutional neural network-based Intrusion Detection System (IDS) implemented as an SDN application and a Blockchain-based system (BS) to empower application layer and network layer security, respectively.
A significant advantage of the proposed method lies in jointly minimizing the impact of attacks such as command injection and rule injection on SDN-based IIoT architecture layers.
The proposed IDS exhibits superior classification accuracy in both binary and multiclass categories.
\end{abstract}

\section{Introduction}
The Internet of Things (IoT) has been increasingly adopted across various fields, including agriculture, manufacturing, and industry \cite{iot-chen}. The Industrial Internet of Things (IIoT) is an application of IoT in manufacturing and industry that aims to automate industrial processes and achieve effective and appropriate products through data exchange and digitization \cite{iot-qize}. IIoT offers advantages over traditional Supervisory Control and Data Acquisition (SCADA) systems, such as productivity, scalability, and data analysis \cite{test2}, but the increase in connected devices and the lack of security design in older control systems make factories vulnerable to cyber-attacks. Industrial Control Systems (ICS), which include SCADA, Remote Terminal Unit (RTU), and Programmable Logic Controllers (PLCs), play a crucial role in various industrial infrastructures such as nuclear technology. 

IIoT security research has increased due to the inadequacy of conventional methods such as access control mechanisms, firewalls, and encryption against multiple attacks such as denial of service (DoS) \cite{test3}. Machine learning methods have been used to detect intrusions, but they often fail to detect unknown and new security threats due to poor feature selection and classification \cite{test3, comst-ml}. This approach is also not scalable for large volumes of data, and incorrect classification can have catastrophic consequences \cite{test4} such as a nuclear disaster in a nuclear power plant.

Ensuring accurate classification of attacks is crucial for timely analysis \cite{test4, zsmme3}. Unbalanced datasets from ICS, which are usually in a stable and normal state, may affect machine learning algorithms \cite{test4}. To detect attacks quickly and accurately, Deep Learning (DL) methods \cite{test3}, \cite{test4} have shown advantages over traditional machine learning methods, such as the ability to learn features from original data and manage high-dimensional data to extract valuable patterns. Leveraging the potential of DL, this study aims to eliminate the aforementioned limitations by implementing anomaly detection and attack classification using a DL-based IDS.

Besides security, current IIoT architectures suffer from issues such as scalability, monitoring, data management, and flexibility (since operators must manually configure devices whenever a device update request is received) \cite{test5}. Integration of SDN into IIoT architecture addresses these challenges \cite{test5}. In fact, SDN technology decouples the data plane and control plane, allowing for centrally and intelligently control of network behavior \cite{test31}. In addition, SDN programmability enables the integration of advanced services for managing the network, including its security.
Driven by its benefits, this study uses SDN for two primary reasons. Firstly, it leverages its programmability feature to facilitate the timely detection and mitigation of attacks through the IDS; which is implemented as an SDN application. Secondly, it facilitates monitoring, data management, and flexibility in IIoT architectures. 

Nevertheless, besides its advantages, SDN architecture is also prone to several vulnerabilities that can be exploited by attackers, such as man-in-the-middle (MITM) attacks\cite{test6}. Therefore, strengthening SDN security is crucial to reap its benefits in creating a reliable communication infrastructure for IIoT environments. 
Blockchain technology is a promising candidate to meet this goal, 
owing to its inherent features of transparency, immutability, traceability, and decentralization \cite{test32}. It operates as a distributed ledger, which is immutable and tamper-proof, and stores data on a peer-to-peer network.

Extensive research work has been engaged, investigating the security challenges in both SDN and IIoT technologies (e.g., \cite{test3},\cite{test4},\cite{test22},\cite{test23}) as well as leveraging SDN in IIoT for its perceived benefits (e.g.,\cite{test5},\cite{test14}). However, none of existing contributions has comprehensively addressed security concerns in both technologies in an integrated way.  
Specifically, there is a lack of research that addresses security issues holistically, rather than separately, with the aim of mitigating the impact of attacks on the various layers of SDN-based IIoT architecture simultaneously. 

It is important to clarify that our objective is not to propose a method that is superior to previous research in terms of IDS accuracy or Blockchain overhead. Rather, the primary differentiator of our approach from related work lies in the proposal of a novel and improved integrated method that aims to minimize the effects of IIoT security attacks and SDN security attacks on the various layers of an SDN-based IIoT architecture. This is achieved through the use of two security components -- a DL-based IDS and a Blockchain-based system (BS) -- that are integrated in a manner that allows them to work together and complement each other, resulting in a more comprehensive security approach. As a DL algorithm, a convolutional neural network (CNN) is used. Because the selected dataset (version 3) \cite{test25} is supervised and classification-oriented, CNN has been chosen for its capacity to automatically extract relevant features. The results of our evaluation also showed that CNN outperformed some machine learning algorithms in detecting and classifying attacks. This has been the motivation for pursuing the following:

\begin{itemize}
    \item Introducing a SDN-based IIoT system that combines the benefits of both technologies to improve flexibility and scalability while also addressing security concerns.
    \item Proposing a novel method that utilizes two security components. Through this improved approach, the impact of security attacks across multiple layers of the SDN-based IIoT architecture will be minimized.
    \item Presenting and implementing a system model and attack model about SCADA network attacks and attacks specific to SDN networks.
    \item Evaluating the efficiency of the proposed solution against the attacks outlined in the attack scenario.
\end{itemize}

The rest of this paper is organized as follows. Section \ref{sec:related} categorizes existing articles related to the field. In Section \ref{sec:methodology}, the proposed method, system model, and intended attack model are presented. Sections \ref{sec:implementation} and \ref{sec:evaluation} present implementation details and evaluation results, respectively. Finally, Section \ref{sec:conclusion} concludes this paper and highlights some of our future work in this area.

\section{Related Work} \label{sec:related}
Three categories of related research are reviewed in this section, namely: IoT/IIoT Security and IDS-based solutions, SDN-based IoT/IIoT Security, and SDN Security and Blockchain-based solutions.

\subsection{ IoT/IIoT Security and IDS-based Solutions}
Balil \textit{et al.} \cite{test2} discussed classification and mitigation solutions for IIoT attacks. In \cite{test3}, a two-step detection system is proposed: a machine learning-based anomaly detection module is used in the first step for binary classification. The output of the first step is used as the input for the CNN-Long short-term memory (LSTM) algorithm for multiclass classification in the second step. A 2D CNN algorithm is introduced in \cite{test4} to identify anomalies in industrial traffic. The idea is to convert one-dimensional traffic data into two-dimensional images representing specific traffic classes. As a result, CNN is capable of classifying these images effectively in order to detect anomalies. Rakas \textit{et al.} \cite{test11} reviewed recent articles on SCADA security using IDS, and Alotaibi \textit{et al.} \cite{test12} used stacked deep learning to detect malicious attacks targeting IoT devices in smart homes and smart grids. Gao \cite{test30} designed a SCADA anomaly-based IDS with different algorithms, such as Decision Tree (DT).

\subsection{SDN-based IoT/IIoT Security}
The use of SDN and blockchain in IIoT is proposed to enhance smart grid flexibility, energy optimization, and security against attacks in various articles, including \cite{test5}, and \cite{test14}. In \cite{test5}, SDN detects vulnerabilities and attacks and facilitates continuous IoT device monitoring. The authors leveraged Blockchain to secure Cloud data, counter Distributed DoS (DDoS) attack threats, and guard against distributed controller attacks. Machine learning-based models and clustering algorithms are also suggested to optimize resource consumption and prevent attacks. SDN technology is used in \cite{test16} and \cite{test17} to prevent MITM and DDoS attacks and enhance scalability and flexibility in IoT networks. Haseeb \textit{et al.} \cite{test15} proposed a SDN-enabled security model using machine learning to improve network consumption and delivery of the Internet of Medical Things (IoMT) services on time. SDN clusters the nodes and optimizes routing performance using the unsupervised learning algorithm.  Moreover, the intelligent centralized SDN controller protects data, minimizes power consumption, and manages critical infrastructure effectively, safeguarding against malicious users and unauthorized requests.

\subsection{ SDN Security and Blockchain-based Solutions}
Liu \textit{et al.} \cite{test6} reviewed different types of attacks on SDN. Scott \textit{et al.} \cite{test20} investigated vulnerabilities in software-based networks and suggested security solutions. Derhab \textit{et al.} \cite{test22} used a distributed controller and a blockchain between controllers to prevent flow rule injection attacks, and Boss \textit{et al.} \cite{test23} prevent DDoS attacks using blockchain technology.

A number of articles investigated the issue of SDN security and IIoT networks, emphasizing the demand for an integrated solution to ensure network security and avoid additional costs. The work in \cite{test14} introduces a Random Subspace Learning (RSL)-based IDS method for IIoT attacks and a blockchain-based method for SDN attacks, but does not demonstrate how to implement detection based on the blockchain, and the role of SDN is not clearly defined. 
In contrast, our work outlines a novel approach for detecting SDN attacks based on the southern interface and emphasizes the role of SDN as a key component in detecting IIoT-related attacks through the use of an IDS. We make use of SDN's programmability feature to achieve this goal.


\section{Methodology} \label{sec:methodology}
This section describes the proposed approach, the attack model, and the system model of the desired architecture.

\subsection{System Model}
\begin{figure}[]
\centering
\resizebox{3.3in}{!}{
\includegraphics{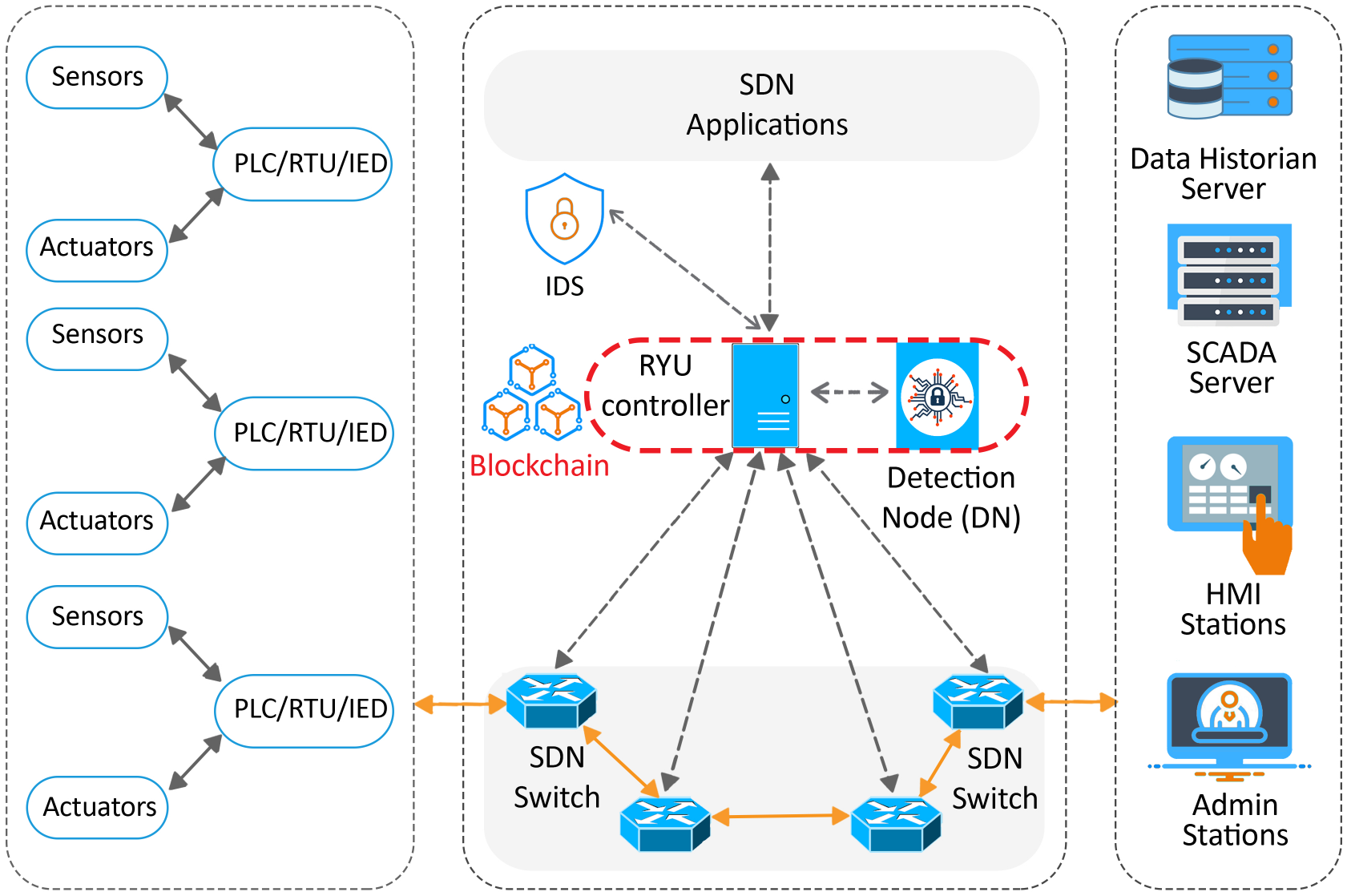}
}
  \caption{Layered architecture of SDN-based IIoT networks.}
    \label{figure:system_model}
    \vspace{-0.4cm}
\end{figure}

The proposed system model, illustrated in Fig. \ref{figure:system_model}, uses SDN technology at the network layer to transmit environmental information from sensors to industrial controllers and ultimately to the control room. The SDN-based network is responsible for transmitting information and making decisions based on data received from the sensors. Two security components, namely blockchain and IDS, are located within the system to enhance its security.

Fig. \ref{figure:system_model} shows the blockchain consisting of two nodes; an SDN controller/a block generator and a Detection Node (DN). The block generator has read and write access, while the DN can only read the block.  In addition, the IDS is also placed as an application on the SDN controller - the IDS is trained by CNN algorithm. The blockchain plays a crucial role in our work by supporting the DN in more accurately detecting attacks.

\subsection{Attack Model}
\begin{figure}[]
\centering
\resizebox{\linewidth}{!}{
\includegraphics[width=3.5in]{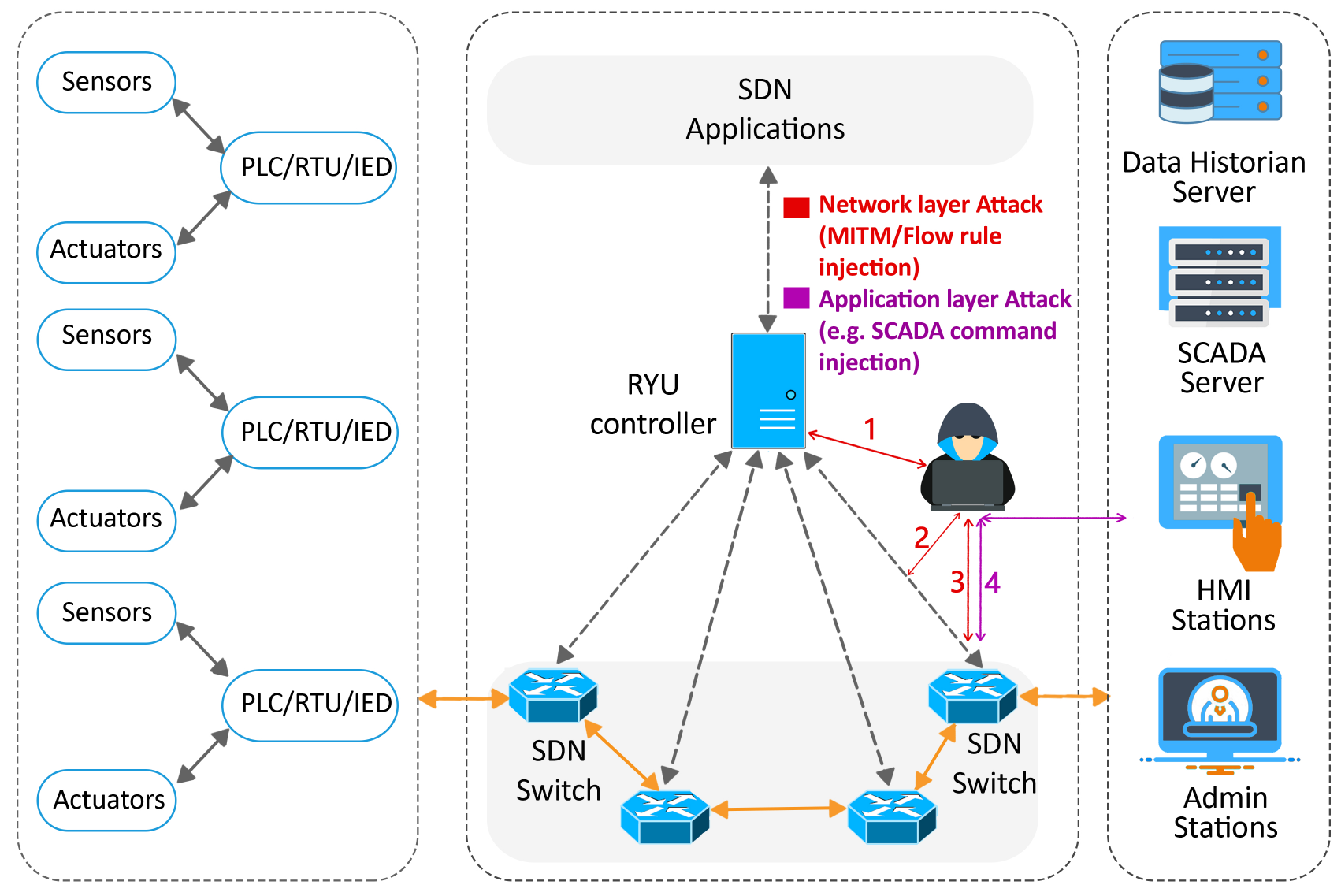}
}
  \caption{Attack model.}
    \label{figure:attack_model}
    \vspace{-0.5cm}
\end{figure}

Fig. \ref{figure:attack_model} depicts a proposed attack model considering MITM attacks at the network layer, involving the injection of flow rules into switches' flow tables and command injection attacks at the application layer (the control room). Two types of attacks are considered: purple attacks modifying packet payload and red attacks changing packet header. Purple attacks alter actuator performance, while red attacks redirect packets to unintended destinations, causing chaos in the network. The control room's command could reach the wrong actuator and unintentionally perform an undesired action due to such attacks.

\subsection{Method Suggested}
In this method, both the packet payload and header are examined to ensure that neither command injection nor flow rule injection has taken place.

The first assumption is that a packet enters the switch. The switch performs an action on a packet if it is defined in its flow table, otherwise it sends it to the controller. Before considering any action, the controller forwards the packet to the IDS for detection of the malicious payload.  If the IDS detects a malicious payload, the controller blocks the attacker, but if the packet payload is safe, the header is checked by the BS which includes a controller node and a DN. Thus, the BS is activated once the IDS has confirmed that the packet payload is safe. 

\begin{figure}[]
\centering
\resizebox{8cm}{!}{
\includegraphics[width=3in]{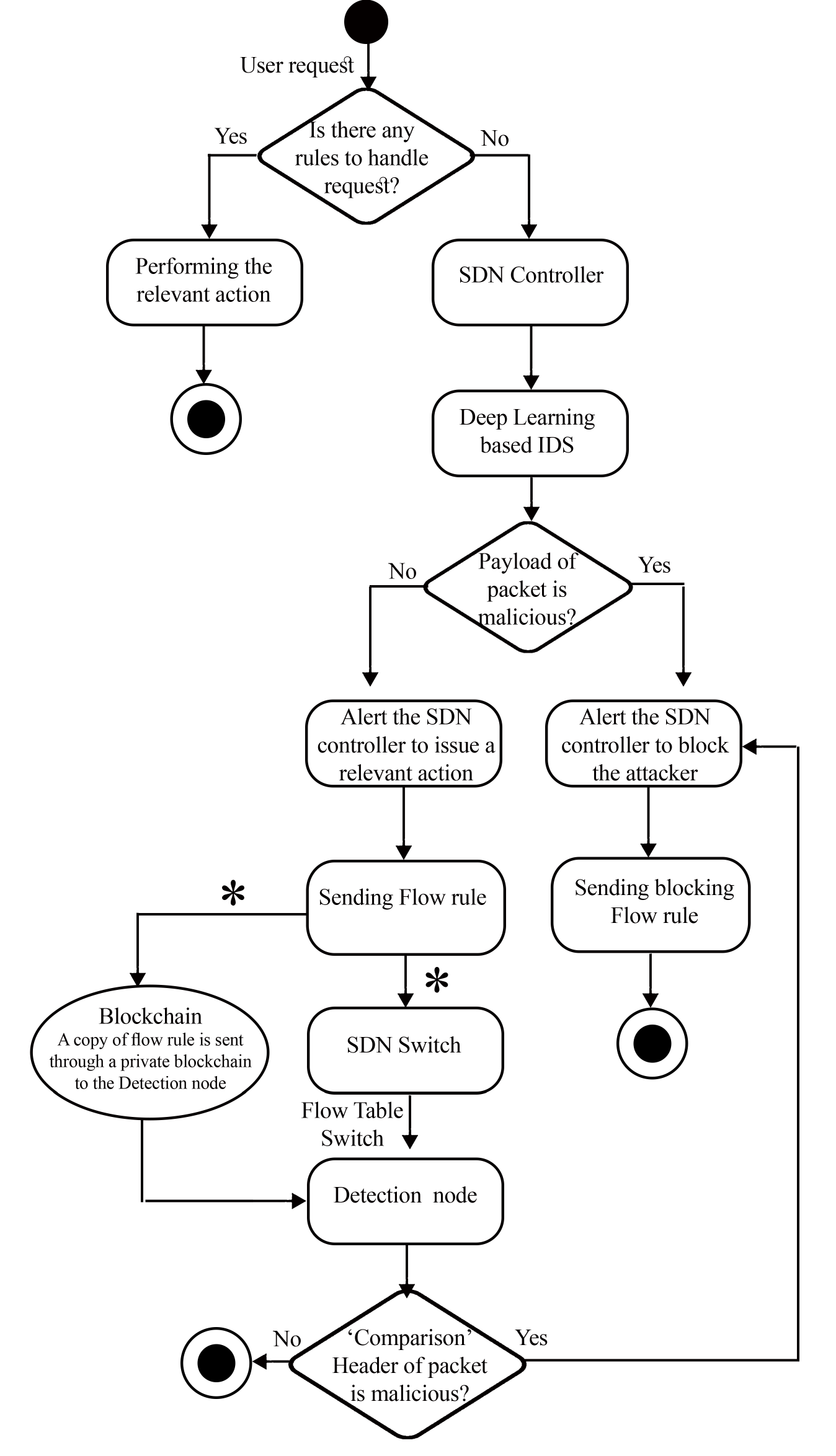}
}
  \caption{Flow chart of the proposed method.   }
    \label{figure:flow_chart}
    \vspace{-0.7cm}
\end{figure}

Due to this, once the IDS has determined that the packet payload is normal, the controller decides on the packet by sending a flow rule to the switches in order to update their flow tables. Simultaneously, the blockchain is used to transmit the same flow rule to the DN. The DN stores the flow rules received from the controller via the blockchain in a file and requests the switch logs for analysis.  Following that, DN saves the switch logs in another file. As a result, DN compares these two files to ensure that the flow rule has not been changed along the way (i.e., southern interface) due to a possible MITM attack.  When the DN detects a change in the values sent by comparing these two files, it will generate an attack warning for the controller, otherwise, it will generate a warning for safe flow for the controller (the packet header is also safe). Consequently, by sending a malicious packet payload or a false flow rule, the attacker is not able to reach his target in this case. Therefore, neither the command sent from the controller room to change the performance of the actuators nor the packet headers to get the packet to the wrong destination have changed. A description of the proposed method is provided in Fig.\ref{figure:flow_chart}. In Fig.\ref{figure:flow_chart}, the arrows specified in the “Sending flow rules” section with the “*” sign occur simultaneously. 

\section{Implementation} \label{sec:implementation}
The proposed method is implemented using the Python programming language and the Numpy 1.18.5, Pandas 3.8.10, Keras1.1.2, and Sklearn 0.22.2 libraries to implement the IDS based on the CNN algorithm. Additionally, the MultiChain private blockchain is implemented using the Savoir library. Mininet 2.3.0, OpenFlow 1.3, and Ryu Controller are used as simulators, southbound interface, and SDN controller, respectively.
In addition, two Ubuntu 20.04 virtual machines are used and the specifications of the systems and programs that are installed on them are listed in Table \ref{table:specification}.

This study uses a dataset in the field of natural gas pipelines \cite{test25}. The dataset contains 27 features. It includes 8 classifications, one of which is designed for normal mode and seven for the attack. The specific classifications of the natural gas pipeline dataset are shown in Table \ref{table:attackclass_f1}. More details on the description and classification of these features are in \cite{test26}.

The CNN-based IDS is implemented after performing data pre-processing operations to clean the input data and improve accuracy. Features with only one value are removed, reducing the number of features from 27 to 18. Data grouping is used to balance the data in each of the eight classifications, resulting in four classifications: normal, injection attack, reconnaissance attack, and DoS attack. This is achieved by grouping five types of injection attacks into one attack group and defining one type of attack as an injection attack. It is worth mentioning that the DoS attack detected by the IDS in this study is a false state injection, not a flooding attack.

The dataset is split into three parts: 70\% for training, 15\% for validation, and 15\% for testing. By using the MinMaxScalar, the data is normalized between 0 and 1.



\begin{table}[t!]
\caption{Attack classification in the data set.}
\vspace{-0.45cm}
\label{table:attackclass_f1}
\begin{center}
\begin{tabular}{|c|c|c|c|}
\hline

Abbreviation & Label & Label type  \\ \hline
Normal & 0 & Normal Behaviour \\ \hline
NMRI & 1& Naive Malicious Response Injection \\ \hline
CMRI& 2 & Complex Malicious Response Injection \\ \hline
MSCI & 3 & Malicious State Command Injection \\ \hline
MPCI & 4 & Malicious Parameter Command Injection  \\ \hline
MFCI& 5& Malicious Function Code Injection \\ \hline
DoS & 6& Denial of Service  \\ \hline
Recon & 7& Reconnaissance \\ \hline

\end{tabular}
\end{center}
\vspace{-0.5cm}
\end{table}

\begin{table}[t!]
\caption{Specifications of systems used in research.}
\vspace{-0.45cm}
\label{table:specification}
\begin{center}
\begin{tabular}{|c|c|c|c|}
\hline
System & Operating System & RAM & Installed Program \\ \hline
1 & Ubuntu 20.04 & 8 GB & Mininet 2.3.0, Ryu 4.34 \\ \hline
2& Ubuntu 20.04 & 4 GB & Multichain 2.2 \\

\hline
\end{tabular}
\end{center}
\vspace{-0.7cm}
\end{table}

Afterward, the CNN model is defined. It has 2 layers of 1D-convolution, 1 layer of Max-pooling, 1 layer of Average-pooling, and 2 Fully connected layers. The filter size in the convolution layer is 3, while in the max-pooling layer, it is 2. The Relu activation function is used in the convolution layer, sigmoid for binary classification, and softmax for multi-class classification in the last layer. The batch size is set at 100, with SGD as the optimization function. The learning rate and momentum are set as 0.01 and 0.8, respectively.

It is important to note that the selection of hyperparameters, including the number and type of layers, optimizer function, and learning rate, were determined through a process of trial and error. We tested various learning rates of 0.1 and 0.01 as well as momentum values of 0.8, 0.9, and 0.99. Ultimately, we found that the combination of a learning rate of 0.01 and momentum of 0.8 by SGD produced the most optimal results.
However, due to space constraints, in this paper, we are only able to present the results of the model trained with SGD using a learning rate of 0.01 and momentum of 0.8. 

Following the initial steps and defining the model using the aforementioned values, the model is implemented with 30 training epochs and then evaluated according to metrics. Section \ref{sec:evaluation} presents the results of this evaluation based on the considered metrics. Following the implementation of IDS, each of the systems listed in Table \ref{table:specification} will be described.

\subsection{System 1 Operation in Implementation}
The system is simulated using Mininet and Ryu controller, with a network topology consisting of 2 switches, 3 hosts, and 1 controller with communication links. The presented topology uses a host called Client for legal communication, a host called Attacker for sending the false flow rules, and a host called Server for receiving requests from users. After implementing the topology in Mininet and its connection to the controller, it is possible to send packets with a specific payload. The trained CNN-based IDS model is saved in a file and called on the Ryu controller. An IDS Python application is created to receive packet payload, read its specifications, call the corresponding IDS, define flow rules sent by the controller, and perform blockchain operations. 

\begin{figure}[]\centering
\includegraphics[width=3.5in]{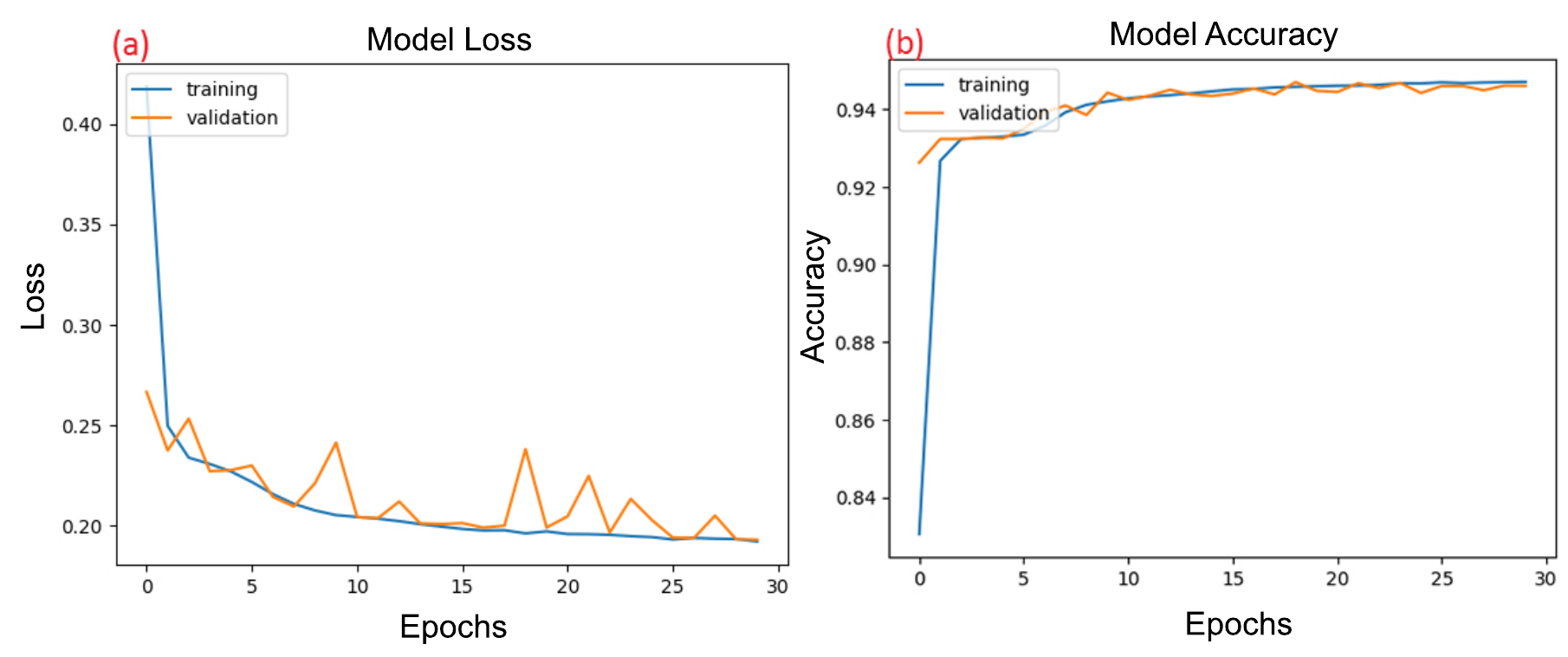}
  \caption{Model's accuracy and loss at every epoch in binary-class classification.}
    \label{figure:accuracy_loss_binary}
    \vspace{-0.4cm}
\end{figure}
\begin{figure}[]\centering
\includegraphics[width=3.5in]{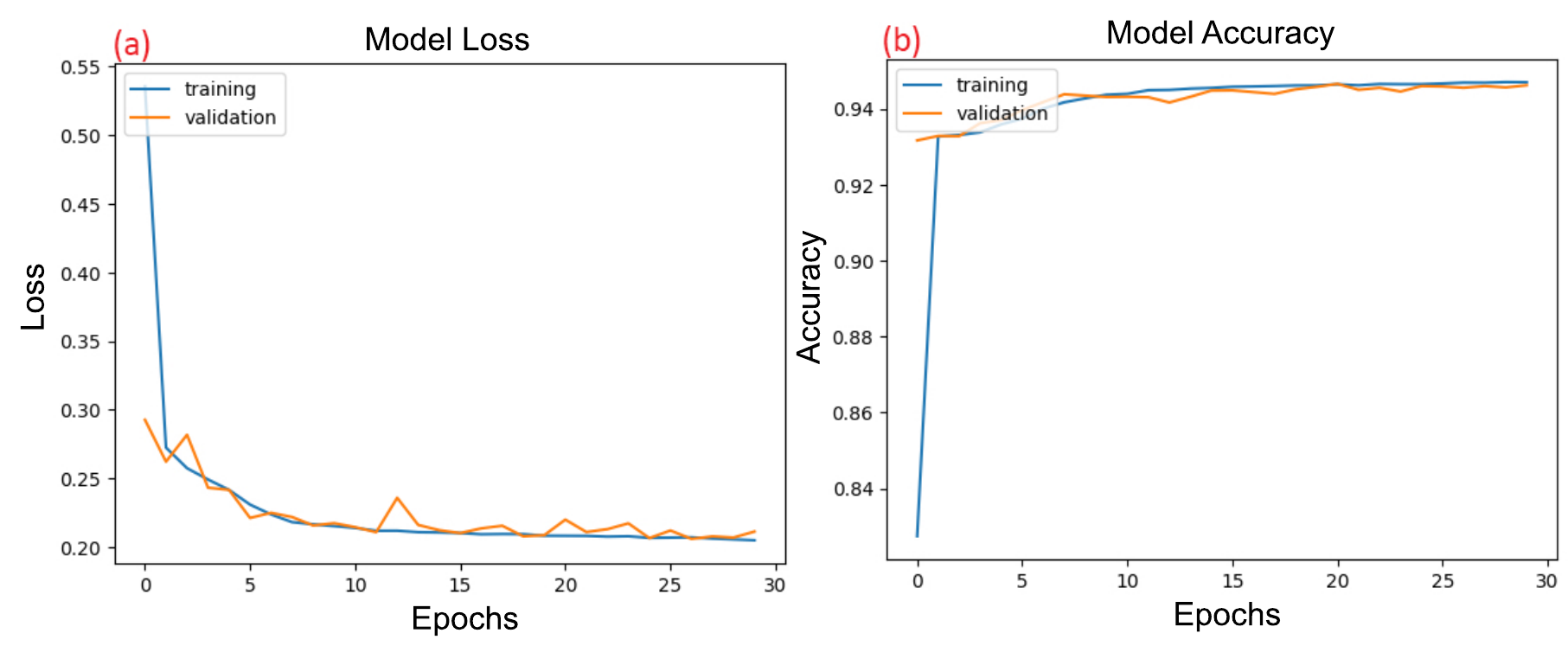}
\vspace{-0.5cm}
  \caption{Model's accuracy and loss at every epoch in multi-class classification.}
    \label{figure:accuracy_loss_multi}
    \vspace{-0.5cm}
\end{figure}

\subsection{System 2 Operation in Implementation}
The controller sends a flow rule to the switch after the client hosts send a normal payload and the IDS verifies it. The DN (system 2) receives a copy of the same flow rule via Savoir, which is transmitted to the switch. The DN compares the switch flow table with the flow rule received from the blockchain. A MITM occurs when the number of flow table rows received from the switch does not match the number of flow table rows received from the controller. If the rows match in both files but the content does not, it suggests a modification in the flow rule at the switch. This is reflected in the "Hard\_age" feature of the switch flow table, which records the time since the entry was last modified. Hence, Hard\_age can be used to detect flow table changes. Whenever a flow rule injection attack is detected, the DN notifies the controller that the packet is malicious, indicating that the flow rule sent from the SDN controller to the switch has been tampered with.

\section{Evaluation} \label{sec:evaluation}
This section evaluates the IDS before calling in the Ryu controller, and in the next step, we evaluate the IDS after calling in the Ryu controller and the BS.

\subsection{IDS Evaluation Before Calling in Ryu}
The proposed IDS is evaluated using four performance metrics: accuracy, precision, recall, and F1-score. 

The accuracy represents the overall performance of the classifier.  Precision and recall ensure that the results were not distorted by too many normal samples (unbalanced dataset). Finally, F1-score acted as a reconciler between precision and recall. 


\begin{table}[t!]
\caption{Binary classification performance results.}
\vspace{-0.45cm}
\label{table:evaluation_a_p_r}
\begin{center}
\begin{tabular}{|c|c|c|c|}
\hline
Class & Precision & Recall & F1 \\ \hline
Normal & 94.17\% &	98.3\% & 96.6\% \\ \hline
Attack & 96.39\% & 89.65\% & 92.90\%  \\ \hline
Average results & 95.28\% & 93.84\% & 94.48\% \\ \hline
\end{tabular}
\end{center}
\vspace{-0.5cm}
\end{table}

\begin{table}[t!]
\caption{Multi-class classification performance results.}
\vspace{-0.45cm}
\label{table:evaluation_f1}
\begin{center}
\begin{tabular}{|c|c|c|c|}
\hline

Class & Precision & Recall & F1	 \\ \hline
Normal & 94.11\% & 98.04\% & 96.3\%	\\ \hline
Injection & 95.32\% & 88.29\% & 91.67\% \\ \hline
DoS & 98.67\% & 69.13\% & 81.30\%  \\ \hline
Reconnaissance & 100\% & 100\% & 100\% \\ \hline
Average results & 97.02\% & 88.86\% & 92.31\% \\ \hline

\end{tabular}
\end{center}
\vspace{-0.7cm}
\end{table}

Fig. \ref{figure:accuracy_loss_binary}, Fig. \ref{figure:accuracy_loss_multi}, Table \ref{table:evaluation_a_p_r} and Table \ref{table:evaluation_f1} present the evaluation results of the proposed algorithm based on the above-mentioned metrics. 
Fig. \ref{figure:accuracy_loss_binary} as well as Fig. \ref{figure:accuracy_loss_multi} illustrate the loss (a) and accuracy (b) of the binary-class and multi-class classification models during training, respectively.
An evaluation of the model was done with a learning rate of 0.01 and momentum of 0.8 in 30 epochs.

The accuracy of the model was tested using various hyperparameters and optimization functions. The accuracy of the model in the binary-class mode reached 93.50\% with the learning rate of 0.1 and the momentum of 0.8. With the learning rate of 0.01, it reached 94.75\%. In the multiclass mode, the model accuracy reached 93.30\% with the learning rate of 0.1 and the momentum of 0.8. It reached 94.65\% with the learning rate of 0.01. Model accuracy in the binary-class and multi-class modes reached 63\% when the learning rate was set to 0.1 and the momentum was set to 0.99.

Based on the evaluations, it is demonstrated that the model trained using a learning rate of 0.01 and momentum of 0.8 by SGD outperforms other models.
Table \ref{table:evaluation_a_p_r} shows the outcomes of binary classification, while Table \ref{table:evaluation_f1} demonstrates the results of multi-class classification.

To assess the effectiveness of the proposed algorithm, in comparison to the algorithms presented in \cite{test14} and \cite{test30}, we utilize DT method and the RSL-K-Nearest Neighbor (KNN) approach to evaluate their accuracy performance in detecting SCADA attacks. Table \ref{table:comparison} presents the outcomes of binary and multi-class classification based on the accuracy metric. We experimented RSL-KNN \cite{test14} with different K -- the number of nearest neighbors -- values of 5 and 10, but the best accuracy achieved was below 91.9\% in both binary and multi-class classification. The accuracy of the DT method in both binary and multi-class classification was 92.3\%, as well. We can observe that CNN outperforms  RSL-KNN and DT  under both classification tasks.

\subsection{IDS Evaluation in the Form of an SDN Application}
After ensuring that the controller is properly prepared, packets containing the desired payloads are sent through the server. The effectiveness of the IDS application in detecting both malicious and normal payloads is then assessed through careful evaluation. The attacker system sends a request with a malicious payload to the server, which is then sent to the controller for a decision as there is no appropriate action in the switch flow table. The controller forwards the packet to the IDS application, which announces a value of 1 if the packet contains a detected malicious payload. The IDS application then sends a command to the controller to block the source system. 

\begin{table}[t!]
\caption{Algorithm comparison results.}
\vspace{-0.45cm}
\label{table:comparison}
\begin{center}
\begin{tabular}{|c|c|c|}
\hline

\multirow{2}{*}{Algorithms} & Accuracy in binary  & Accuracy in multi-class \\ 
& class classification & classification \\ \hline
CNN & 94.75\% & 94.65\% \\ \hline
\multirow{2}{*}{RSL-KNN \cite{test14}} & (K = 10) 90\% & (K = 10) 90.2\% \\

& (K = 5) 91.9\% & (K = 5) 91.9\% \\ \hline
DT \cite{test30} & 92.3\% & 92.3\%  \\ \hline

\end{tabular}
\end{center}
\vspace{-0.4cm}
\end{table}



    After that, the IDS application is tested by sending normal payloads to the server. The flow rule between the client and server is sent to the switches via the SDN controller, and a copy of the controller flow rule sent to the switch is collected. The collected flow tables are sent to the DN (system 2) using the Savoir module as a block. The DN receives the block and stores the results in a file. The DN then saves the content of the switch flow table in another file. The existence of different number of rows in flow tables in the two files when compared by DN indicates a MITM attack. The system also reports a modification when the switch flow table contains Hard\_age and the number of rows is equal. 




\section{Conclusion} \label{sec:conclusion}
This research utilizes CNN-based IDS and blockchain as complementary components to enhance the security of SDN-based IIoT architecture. SDN's programmability, centralized controller, and network-wide view make it a crucial factor in ensuring IIoT security. The study achieved 94.75\% accuracy in binary classification and 94.65\% in multi-class classification, along with satisfactory precision, recall, and F1-score metrics. The proposed method effectively detects malicious payloads and prevents their occurrence, as well as detects flow rule injection attacks. Future work could involve using balancing methods and alternative algorithms to improve classification metrics and multi-controllers can be implemented to prevent system bottlenecks. 
\vspace{-0.5cm}
\section*{Acknowledgment}

This research work is partially supported by the Business Finland 6Bridge 6Core project under Grant No. 8410/31/2022, the Research Council of Finland (former Academy of Finland) IDEA-MILL project (Grant No. 352428), the Research Council of Finland 6G Flagship program (Grant No. 346208), and the European Union’s Horizon Europe research and innovation programme HORIZON-JU-SNS-2022 under the RIGOUROUS project (Grant No. 101095933). The paper reflects only the authors' views. The Commission is not responsible for any use that may be made of the information it contains.

\vspace{-0.5cm}


\end{document}